# Resistance Quenching in Graphene Interconnects

Q. Shao, G. Liu, D. Teweldebrhan and A.A. Balandin[*]

Nano-Device Laboratory, Department of Electrical Engineering, University of California – Riverside, Riverside, California 92521 USA

Materials Science and Engineering Program, Bourns College of Engineering, University of California – Riverside, Riverside, California 92521 USA

**Abstract**

We investigated experimentally the high-temperature electrical resistance of graphene interconnects. The test structures were fabricated using the focused ion beam from the single and bi-layer graphene produced by mechanical exfoliation. It was found that as temperature increases from 300 to 500K the resistance of the single- and bi-layer graphene interconnects drops down by 30% and 70%, respectively. The quenching and temperature dependence of the resistance were explained by the thermal generation of the electron-hole pairs and acoustic phonon scattering. The obtained results are important for the proposed applications of graphene as interconnects in integrated circuits.

---

[*] Corresponding author; E-mail address: balandin@ee.ucr.edu ; web-address: http://ndl.ee.ucr.edu

As the electronic industry aggressively moves towards nanometer designs thermal issues are becoming increasingly important for the high-end electronic chips. The integrated circuit (IC) performance is now limited by the maximum power, which can be dissipated without exceeding the maximum junction temperature setup by the reliability requirements [1-2]. According to the International Technology Roadmap for Semiconductors (ITRS) projections the volumetric heat generation rates within interconnects will be approaching $P=j^2 r \sim 3.3 \times 10^4$ W/mm$^3$ assuming a current density $j=3.9$ MA/cm$^2$ and a resistivity $r=2.2$ $\mu\Omega$cm. The self-heating problem is aggravated by the increased integration densities, faster clock speed, high dissipation power density in interconnect networks, increased total thermal boundary resistance of the chip layers, incorporation of the alternative dielectric materials with low thermal conductivity values as well as acoustic phonon confinement effects in nanometer scale structures [3-4].

One of the approaches to mitigate the self-heating problem is to incorporate into the chip interconnect design the materials with low electrical resistance and high thermal conductivity. Carbon nanotubes (CNTs) have been considered for interconnects in the very large scale integrated (VLSI) circuit applications [5-6]. Graphene, a form of carbon consisting of separate atomic planes of sp$^2$-bound atoms [7], was also proposed for the interconnect applications [8-9]. Graphene manifests extremely high room temperature (RT) electron mobility as high as $\sim 15000$ cm$^2$V$^{-1}$s$^{-1}$. It was recently discovered by Balandin *et al* [10-11] that graphene is also a superior heat conductor with the RT thermal conductivity in the range of 3100 – 5300 W/mK [10-11]. The latter adds validity to the proposed interconnect applications of graphene owing to potential benefit for thermal management. In this case, graphene interconnects may be used for high-heat flux cooling and help with lateral heat spreading and hot-spot removal. Since conventional VLSI circuits operate at elevated temperatures (100-200K above RT) it is important to understand how the electrical resistance of graphene interconnects changes as the temperature increases from 300 to 500K.

In this letter we show that the electrical resistance of graphene, which is a semi-metal with zero band-gap [7], undergoes strong quenching as the temperature exceeds RT. Interestingly, this behavior is opposite of that manifested by some technologically important bulk semimetals such as bismuth telluride or related alloys widely used in thermoelectrics [12]. We have produced a large number of graphene layers by mechanical exfoliation from the bulk highly oriented pyrolitic graphite (HOPG) and from the high-pressure high-temperature (HPHT) grown material[13]. The single-layer graphene (SLG) and bi-layer graphene (BLG) were found with the help of micro-Raman spectroscopy through the 2D-band deconvolution procedure [14-16].

Raman spectra were measured at RT using Renishaw instrument under 488 nm excitation wavelength in the backscattering configuration [15-16]. Fig. 1 shows characteristic Raman spectrum with clearly distinguishable *G* peak and *2D* band. The position of *G* peak and shape of *2D* band confirm that the examined flake is SLG. The disorder-induced *D* peak is absent in the scattering spectra from HPHT graphene (its expected position is indicated by an arrow), which suggests a high quality of SLG material. Graphene layers have been transferred to Si substrates with the electrically insulating oxide films of thickness $W=0.3$ µm grown on top of them. A set of SLG and BLG resistors contacted by platinum (Pt) electrodes have been fabricated using Leo XB1540 Focused Ion Beam (FIB) system. The absence of leakage current through the oxide layer was verified by applying very high bias (up to ~20 V) between the top electrodes and back gate (metallization on the back side of the Si substrate) and ensuring the resulting current is negligibly small.

The graphene resistors between two metal electrodes on insulating oxide layer, which can be considered as prototype graphene interconnects, have been electrically characterized in the temperature range T=300 – 500K. The temperature was controlled externally through the Signatone probe-station hot chuck. In Fig. 2 we present typical current-voltage (IV) characteristics for SLG interconnect fabricated from HOPG material. The inset shows a scanning electron microscopy (SEM) image of SLG resistor between two Pt electrodes. The electrical properties of interconnects made of HOPG and HPHT graphene were similar for the examined set of samples. As one can see the resistors are Ohmic and the current increases with increasing temperature. Such a behavior is characteristic for intrinsic semiconductors where the electrical conductivity $\sigma_i$ obeys the following temperature dependence [17] $\sigma_I \sim exp\{-\Delta E_i/(2k_BT)\}$ (here $\Delta E_i$ is the band-gap and $k_B$ is the Boltzmann's constant). The decreasing resistance of semiconductors with *T* is due to growing concentration of the thermally generated electron-hole pairs. It is influenced by the band-gap renormalization and carrier scattering on phonons as the temperature changes [17]. It is interesting to note that the measured trend in graphene is opposite of that in bulk semimetals of bismuth type (e.g. $Bi_xSb_{1-x}$, Bi-Ti, Bi-Sn) where resistivity $r$ follows the law [12] $r=r_o+AT$ (here $A$ is a positive constant between (2.3 – 14) x $10^{-7}$ $\Omega$cm/K). Such dependence for semimetals and metals is explained by the increasing electron – phonon scattering at elevated temperature [18]. In metals the number of charge carriers does not change with temperature but the interaction with phonons increases. The latter results in the temperature dependence of the type $R=R_o[1+a(T-T_o)]$, where $a$ is the temperature coefficient of resistance. At low temperature resistance is limited by impurities, which leads to increasing mobility and decreasing resistance with $T$. The temperature dependence of resistance in bismuth near RT reverses when one makes a nanostructure out of it, e.g. nanowire, with the lateral dimensions below some critical value. In this case a semimetal – semiconductor transition is induced by quantum confinement, which results in the experimentally observed change in the resistance temperature dependence [19].

Fig. 3 presents the electrical resistance for SLG and BLG interconnects as a function of temperature. The resistances were normalized to their values at RT for better comparison. The plot also shows a theoretical curve for the SLG resistor obtained from the model recently proposed by Vasko and Ryzhii [20] and re-normalized to RT value for better comparison. Our experimentally obtained dependence for SLG is in excellent agreement with the calculations. According to the theory proposed in Ref. [20] the decrease in resistance at RT and above comes from the thermal generation of carriers while the values and shape of the resistance curve are determined by electron and hole scattering on the long and short range disorder and acoustic phonons. Cheianov and Falko [21] also predicted a negative linear $T$ dependence of resistivity $R(T)$ in graphene described by the expression $R(T)=R(0)-(h/e^2)(4TV_o/hv^2E_Ft_o)$, where $h$ is the Plank's constant, $e$ is the charge of an electron, $E_F$ is the Fermi energy, $t_o$ is a backscattering rate from atomically sharp defects in graphene lattice, which does not include Coulomb scatterers, $v$ is the velocity, and $V_o$ is a characteristic interaction constant [21]. For our samples we obtained the following linear analytical approximation for the high-temperature normalized resistance of SLG: $R(T)/R(T=300K)=1.436 - 0.00147T$. From the known carrier Fermi velocity in graphene of $V_F\sim10^8$ cm/s and the experimentally determined temperature when the resistance quenching sets up (~300 – 350K) we can estimate the correlation length for the disorder scattering in our graphene resistors [20], i.e., $l_c \sim V_F h/(2\pi T k_B)$, to be around 22 – 25 nm. The origin of the difference in the resistance temperature dependence for SLG and BLG requires further theoretical and experimental investigation.

It is illustrative to compare electrical resistance of graphene with that of bulk graphite and other carbon materials. It was known for a long time that single graphite crystals are good electrical conductors along the graphite planes and very poor ones across with the ratio of resistivities above $\sim 10^4$ [22]. There is substantial discrepancy for the reported temperature dependence of the electrical resistance in bulk graphite, which likely can be attributed to the variations in the material quality. From the data presented in Ref. [23-24], the resistance decreases with increasing temperature around RT although in one case the decrease in sub-linear while in another case it is superlinear. The high-temperature resistance decreases with temperature in the coke base carbon (T=300 – 800K) and graphitized lampblack base carbon (T=300 – 2000K) as summarized in Ref. [25] although the dependence is very different from what we have measured for graphene. In some types of carbon, e.g. graphitized coke base carbon, the decreasing trend reverses to increasing resistivity around 400-500K [25].

In conclusion, we experimentally investigated the high-temperature electrical resistance of graphene single and bi-layer conductors. It was found that as the temperature increases from 300 to 500K the resistance of the single- and bi-layer graphene interconnects drops substantially. In this sense, despite being semimetal with zero-band gap, graphene resistors behave more like intrinsic semiconductors. The observed resistance quenching in graphene resistors may have important implications for the proposed applications in interconnects and thermal management. The resistance quenching in the relevant temperature rage (100-200K above RT) by 30-70% may lead to significant reduction in power dissipation.

*Acknowledgements*

This work was supported, in part, by DARPA – SRC through the FCRP Interconnect Focus Center (IFC) and AFOSR through award A9550-08-1-0100 on Electron and Phonon Engineered Nano- and Heterostructures.

Q. Shao, G. Liu, D. Teweldebrhan and A.A. Balandin 2008

**FIGURE CAPTIONS**

**Figure 1:** Raman spectrum of the graphene flake, which was used for the interconnect fabrication. The position of *G* peak and spectral features of *2D* band confirm the number of atomic layers.

**Figure 2:** High-temperature current – voltage characteristics of graphene resistors. Inset show SEM image of the graphene interconnects contacted through FIB-fabricated platinum electrodes.

**Figure 3:** Normalized electrical resistance of SLG and BLG interconnects as a function of temperature. The theoretical prediction for SLG from Ref. [20] is shown for comparison. Note a strong quenching of the resistance at temperatures above RT. Inset shows a close up SEM image of BLG resistors between two electrodes.

Q. Shao, G. Liu, D. Teweldebrhan and A.A. Balandin 2008

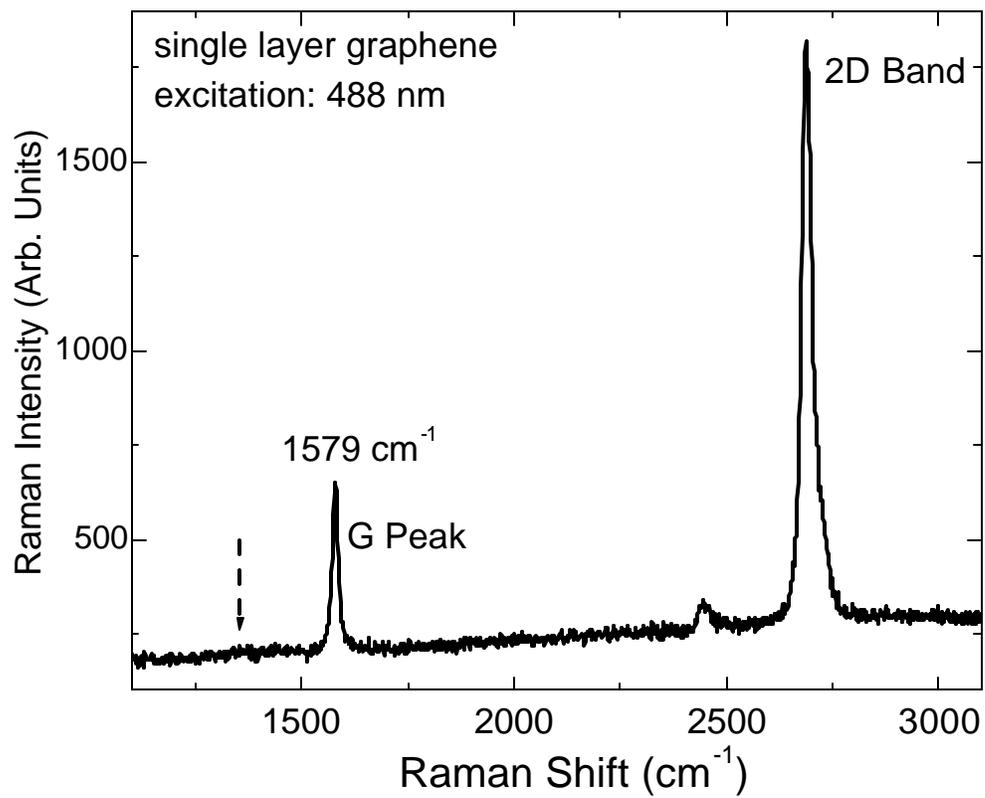

Figure 1: Shao et al.

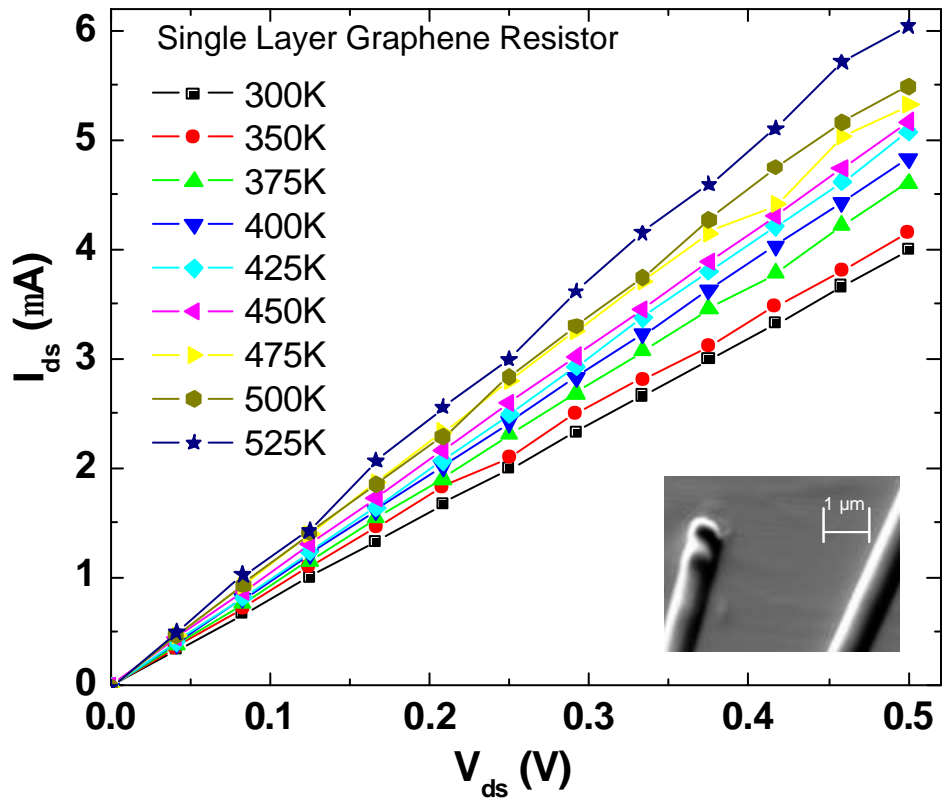

Figure 2: Shao et al.

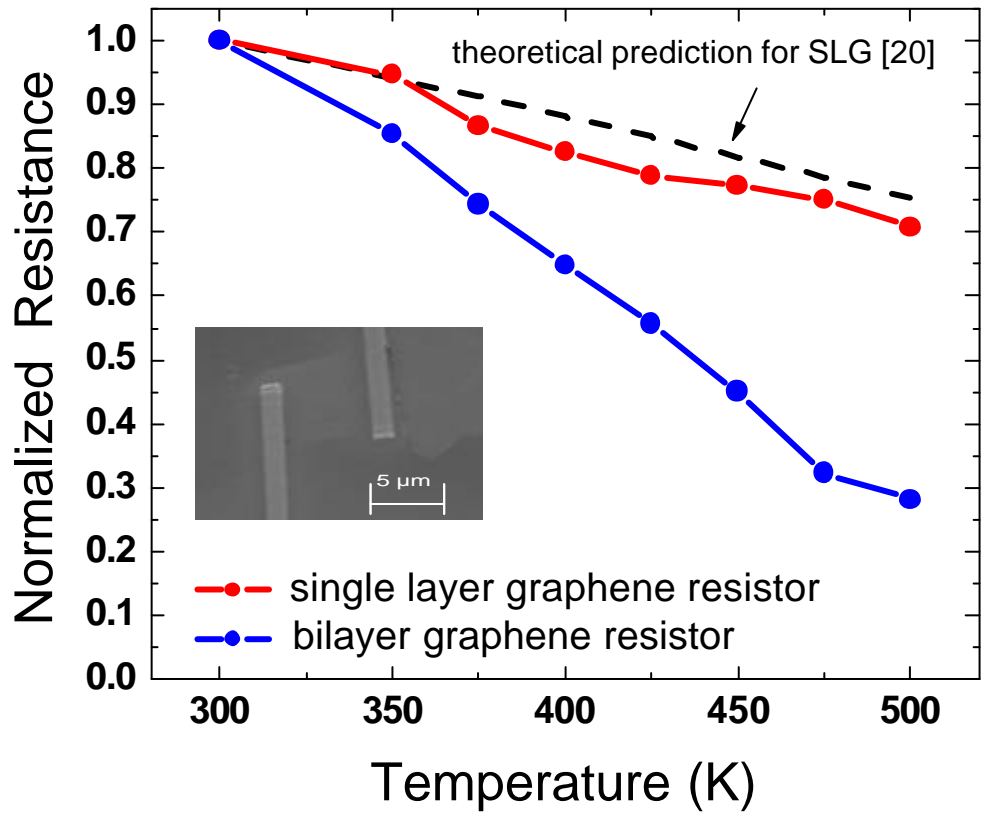

Figure 3: Shao et al.